\documentclass[12pt,doublecolumn]{iopart}
\pdfoutput=1

\usepackage{graphicx}
\usepackage{iopams}

\begin{document}


\title[Diversity of Dynamics in Complex Networks
Without Border Effects]{Characterizing the Diversity of Dynamics in
Complex Networks Without Border Effects}

  \author{Matheus Palhares Viana$^1$, Bruno A. N. Traven\c{c}olo$^1$,
  Esther Tanck$^2$ and Luciano da Fontoura Costa$^{1}$}

  \address{$^1$ Institute of Physics of S\~{a}o Carlos - University of
  S\~{a}o Paulo\\ Av. Trabalhador S\~{a}o Carlense 400, Caixa Postal
  369, CEP 13560-970 \\ S\~{a}o Carlos, S\~ao Paulo, Brazil}
  \address{$^2$ Orthopaedic Research Lab - Radboud University Nijmegen
  Medical Center\\ PO Box 9101, 6500 HB Nijmegen, Phone +31
  243616959\\ The Netherlands} \ead{\mailto{luciano@if.sc.usp.br}}

\begin{abstract}
The importance of structured, complex connectivity patterns found in
several real-world systems is to a great extent related to their
respective effects in constraining and even defining the respective
dynamics.  Yet, while complex networks have been comprehensively
investigated along the last decade in terms of their topological
measurements, relatively less attention has been focused on the
characterization of the respective dynamics.  Introduced recently,
the diversity entropy of complex systems can provide valuable
information about the respective possible unfolding of dynamics.  In
the case of self-avoiding random walks, the situation assumed here,
the diversity measurement allows one to quantify in how many
different places an agent may effectively arrive after a given
number of steps from its initial activity.  Because this measurement
is highly affected by border effects frequently found as a
consequence of network sampling, it becomes critical to devise means
for sound estimation of the diversity without being affected by this
type of artifacts.  We describe such an algorithm and illustrate its
potential with respect to the characterization of the self-avoiding
random walk dynamics in two real-world networks, namely bone canals
and air transportation.
\end{abstract}

\maketitle

\section{INTRODUCTION}

Complex networks (e.g.~\cite{Barabasi2002Survey, Newman2003Survey,
Costa2007Survey}) are natural candidates for representing,
characterizing and modeling a large range of natural and artificial
systems, especially those exhibiting a particularly intricate
(`complex') organization and dynamics.  Indeed, several recent related
works have focused the relationship between the structure and dynamics
of such complex systems (e.g.~\cite{Barabasi2002Survey,
Dorogovtsev2002Survey, Bocaletti2006Survey, Noh2004RandonWalk}). One
particularly important feature to be investigated regards the time
evolution of particular types of dynamics, such as traditional random
walks (e.g.~\cite{Noh2004RandonWalk, Costa2007RandomWalk}), Ising
interaction (e.g.~\cite{Dorogovtsev2002Ising, HERRERO_ISING_PRE_2002,
BIANCONI_ISING_CONDMAT_02}), integration-and-fire
(e.g.~\cite{Costa2008IntegrationFire}), associative memory
(e.g.~\cite{Costa2003Hopfield, Costa2004AssociativeRecall}) among many
other possibilities.  Because of their more purposeful dynamics and
relative simplicity, \emph{self-avoiding walks} stand out as a
particularly interesting type of non-linear dynamics in networks
(e.g.~\cite{Liu2005Dynamics, Costa2008Superedges}).  Instead of
choosing next moves with uniform probability, as in traditional random
walks, self-avoiding walks avoid repeating edges or nodes, therefore
implying in more purposive and effective displacements away from the
starting node.  Moreover, while traditional random walks are
associated to walks, which can be infinite, self-avoiding walks are
intrinsically related to \emph{paths}, which are always of finite size
in finite networks, and therefore more meaningful.

Given a specific dynamics involving a moving agent, such as in
self-avoiding walks, it is interesting to investigate the diversity of
destinations of several walks after starting at specific nodes.  The
importance of such a study stems from the fact that diversity is
immediately related to the influence of the starting nodes on the
other nodes in the network (e.g. diseases can be more effectively
propagated from specific nodes).  One immediate possible means to
quantify the diversity of destinations after $h$ steps along a walk
which started at a given node $i$ would be to identify how many nodes
can be reached at that step $h$ by an infinite number of walks
initiating at $i$.  Though this can be easily determined by
identifying the $h-$th hierarchical (or concentric) neighborhood
(e.g.~\cite{Costa2004PRL}) of node $i$, such a measurement would
provide but a biased indication of the diversity of the walk at that
step.  The problem with this approach is that it does not consider
that the probabilities of getting to the reachable nodes after $h$
steps are typically not uniform because of the heterogeneity of the
network connectivity between the source and destination nodes.  Thus,
it is possible that out of several nodes reachable after $h$ steps,
several of them may be accessed only very sporadically, contributing
little effectively to the overall diversity, while others are reached
frequently.  A more significant means to quantify the diversity of
destinations along the evolution of several types of dynamics in
complex networks was reported recently~\cite{Costa2008Diversity},
consisting in the \emph{diversity entropy} of a node $i$ after $h$
steps.  By calculating the entropy of the transition probabilities to
the reachable nodes, a much more effective quantification of the
diversity of the dynamics can be achieved which also takes into
account the uniformity in which the destination nodes are reached.

Despite the potential of the diversity entropy for revealing important
aspects of the evolution of the dynamics in complex networks, it is
intrinsically plagued by border effects in incomplete networks. For
instance, a transportation network is often constrained to a specific
region of our planet or country, implying in biased calculation of the
diversity entropy at the border nodes located along the regions where
the network was sliced.  The present work reports an effective means
to avoid such border effects.  It is based on an additional important
property of the diversity entropy in revealing the borders of the
network.  More specifically, nodes with low diversity entropy can be
classified as the border of the network. Therefore, by performing some
relatively simple operations described in this article, it becomes
possible to identify a kernel of the original network whose nodes are
unaffected by the borders. It is important to observe that, typically,
one will only want to avoid the borders implied by the selection of a
specific part of the network.  Contrariwise, in the case of a complete
network, its borders are intrinsically important parts which should be
included in the diversity analysis.  We illustrate this procedure by
taking into account two real-world examples: a bone network which was
sliced from a whole pig bone~\cite{Tanck2006Bone}, so that its border
nodes are known a priori, and the US air transportation network, whose
borders are not clearly identifiable.  In the latter case, we consider
the diversity entropy values in order to identify the borders.  In
both cases, the dilation of the borders define a buffer area which
allows the diversity (and other measurements) to be calculated for a
set of kernel nodes withouth being affected by the borders.

This work starts by presenting the basic structural and dynamical
properties of complex networks and follows by describing the proposed
methodology and illustrating it with respect to two real-world
networks.

\section{Structural and Dynamical Properties of Complex Networks}

A complex network is composed by a set of nodes and links connecting
such nodes.  In the case of an unweighted network, the links (edges)
between the nodes can be represented by a matrix $K$, named
\emph{adjacency matrix}, so as that the existence of a link between the nodes
$i$ and $j$ implies $W(i,j)=W(j,i)=1$, with $W(i,j)=W(j,i)=0$ being
otherwise imposed. If two nodes are connected by an edge, they are
said to be \emph{adjacent nodes}. If two edges are associated to the
same node, the are called \emph{adjacent edges}. A sequence of
adjacent edges defines a \emph{walk} over the network. A \emph{path}
is a special type of walk, where no edges or nodes are repeated along
the walk. The \emph{length} of a walk is defined as the number of the
edges of the walk.

An interest way to study the dynamical properties of a network is by
considering a moving agent ``walking'' along its nodes, while guided
by some chosen dynamics. Observe that the moving agent defines a walk
over the network. When the \emph{self-avoiding random walk} dynamics
is chosen, the agent departs from the starting node and choose one of
its neighbors with uniform probability, defining \emph{paths} as they
progress. The others steps are performed in the same manner, but
following the restriction that none of the previously visited nodes or
edges cannot be revisited. The moving agent stops after it reaches a
pre-defined number of steps $H$ or when it can proceed no longer (i.e,
it reaches a terminal node of the network or lacks unvisited neighbors
to proceed).

Let $P_h(i,j)$ be the probability to reach the node $j$ after
departing from the node $i$ through a self-avoiding random walk of
length $h$.  This \emph{transition probability} can be obtained by a
determinist algorithm~\cite{Costa2008Superedges} or by a stochastic
method~\cite{Costa2008Diversity}. The estimation of the transition
probability of the nodes allows to derive a set of complementary
measurements quantifying several aspects of how a node can be reached
and how it can reach other nodes of the network, allowing a
comprehensive characterization of the network
dynamics~\cite{Costa2008Superedges}. In the present work, the
\emph{diversity entropy}~\cite{Costa2008Diversity, Costa2008InOutAcc}
is specifically considered, though the methodology can be easily
adapted to other dynamical measurements. This measurement quantifies
the diversity of the dynamics of the moving agent, taking into account
the probabilities of reaching different nodes after $h$ steps along
the walk.  So, a high diversity entropy value obtained for a given
starting node $i$ after $h$ steps indicates that that node can
effectively access several nodes at that stage of the walk.  It is
important to observe that the diversity entropy provides a more
informative measurement than would be obtained by considering only the
number of distinct nodes accessible after $h$ steps.  This is so
because the diversity entropy considers the dispersion of the
transition probabilities from the starting node $i$ to each of the
reachable nodes after $h$ steps along the walks.

Given a network with $N$ nodes, the \emph{diversity entropy} $E_h$ of
a node $i$ after $h$ steps is given as~\cite{Costa2008Diversity}:

\begin{equation}
    E_h(i) = -\frac{1}{log(N-1)} \sum_{j=1}^{N}
    \left\{
        \begin{array}{rll}
            P_h(i,j)log(P_h(i,j))  & \mbox{if} & P_h(i,j) \neq 0; \\
            0 & \mbox{if} & P_h(i,j) = 0; \\
        \end{array}
    \right.
    \label{eq:diversity}
\end{equation}

Observe that the maximum diversity entropy, equal to one, is reached
when all other $N-1$ nodes of the network can be reached with equal
probability ($1/(N-1)$) from node $i$.  The minimum value of the
diversity entropy (equal to zero) is obtained when no or only one node
is reachable from the starting node $i$.

\begin{figure}[ht]
\begin{center}
    \includegraphics[]{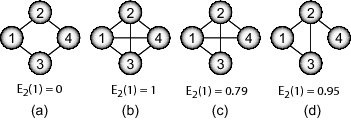}
    \caption{Examples of diversity entropies considering different
    situations.  All cases consider the diversity entropy of node $1$
    after $h=2$ steps (i.e. $E_2(1)$). (a) All paths of length 2 from
    node $1$ end at node $4$, resulting in a null diversity. (b-d) The
    nodes $2$, $3$ and $4$ are reached from node $1$ after two steps,
    increasing the diversity.  (b) Highest diversity case, where the
    access to all other nodes from node $1$ has equal probability. (c)
    The absence of an edge between nodes $3$ and $4$ leads to a
    reduction in the diversity when compared the case (b), mainly
    caused by the high probability of access to node $2$.  (d) The
    absence of the edges $(3,4)$ and $(1,4)$ results in a lower
    diversity value when compared to the case (b), but this reduction
    is not so intense as case (c) because the differences in the
    probability of access to the nodes are relatively smaller.}
    \label{fig:ex_diversity}
\end{center}
\end{figure}

In order to clarify further the concept of diversity, four different
situations are presented in figure~\ref{fig:ex_diversity}. In all
these cases, we consider the diversity entropy for $h=2$ steps after
the moving agents leave from the starting node $1$. In the first
network, shown in (a), only node $4$ is reachable after a
self-avoiding random walk of length two, resulting in a null
diversity. Different situations are implied by the remainder examples
(b-d). In these cases, the nodes $2$, $3$ and $4$ are reached after
two steps from node $1$. For the network depicted in (b), all nodes
are reached with equal probability, resulting in a maximum diversity
($E_2(1) = 1$). The case (c) is similar to (b), except that the link
between the nodes $3$ and $4$ was removed, resulting in a significant
decrease on the diversity ($E_2(1) = 0.79$). In the case (d), not only
the link between nodes $3$ and $4$, but also the link between nodes
$1$ and $4$ were removed. In this case, an interesting result is
observed: the decrease in the diversity to $E_2(1) = 0.95$ was not so
high as in the case (c).  Note that, although the case (c) has one
more connection than (d), the respective diversity is lower because
its additional connection enhances the access to node $2$, unbalancing
the network and diminishing the diversity of access from node $1$ for
$h=2$ (observe that $P_2(1,2) = 2/3$, while
$P_2(1,3)=P_2(1,4)=1/6$). When comparing cases (b) and (d), it is
possible to note that, for both cases, no strong preference of access
to a specific node is found, resulting in a higher value of the
entropy diversity. Indeed, the main difference between these two cases
is that the probabilities of access to the nodes $2$, $3$ and $4$ for
the case (b) are equal, i.e, $P_2(1,2) = P_2(1,3) = P_2(1,4) = 1/3$,
while in (d), $P_2(1,2)= 1/2 > P_2(1,3) = P_2(1,4) = 1/4$.

\section{Diversity-Based Border and Kernel Nodes Detection}

We developed a methodology for avoidance of border effects that
involves the detection of the nodes belonging to the border as well as
the identification of the \emph{kernel nodes}, i.e. nodes whose
specific properties are not influenced by the border
nodes. Particularly, the main purpose of our current work is to find
the nodes whose diversity entropy is not affected by border effects,
so that they can be effectively quantified in terms of that
measurement.

\begin{figure}[ht]
\begin{center}
\includegraphics[width=120mm]{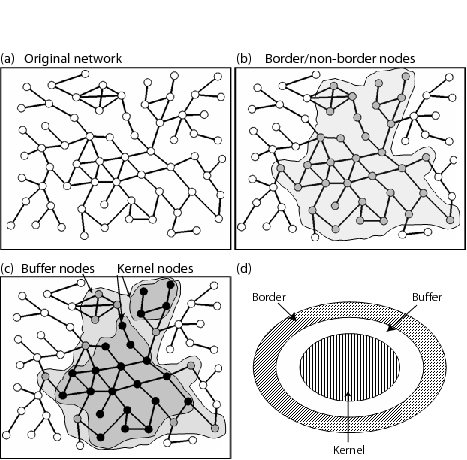}
\caption{Border, buffer and kernel nodes. (a) Original network. (b)
A threshold of the diversity values defines the borders of the network
(white nodes). (c) A dilatation of the border nodes defines the buffer
nodes (in gray) and the kernel nodes (in black). In this example, the
kernel nodes are at a minimum distance of 2 from the border nodes. (d)
A diagram illustrating the relationship between the border,
buffer and kernel nodes.}\label{fig:ex_border_sel}
\end{center}
\end{figure}

The search for these nodes involves the following two main steps: (i)
find the border nodes in the original network; (ii) find the kernel
nodes and take the measurements there. It should be kept in mind that
in the cases where the border nodes are known a-priori, our method
involves only the second step.  Both steps are illustrated in
figure~\ref{fig:ex_border_sel} and described in more detail below.

\textbf{(i) Finding the border nodes}: An interesting property of
the diversity entropy is that the nodes with low diversity values tend
to be located at the borders of the network, in contrast with the
nodes with high diversity, which defines the most central part of the
network. The main idea underling this property is that the self-avoids
walks departing from the nodes near the border have not many choices
except accessing internal nodes. As a consequence, the probability of
accessing the internal nodes becomes higher than the surrounding
border nodes, unbalancing the access distribution and reducing the
node diversity.  So, it becomes possible to identify the border nodes
in terms of the respective diversity entropy values.

Based on this property, the border nodes of the network can be
identified by performing a threshold on the diversity value at a
specific step, or based on the mean value considering all the steps.
Figure~\ref{fig:ex_border_sel}(b) illustrates such a procedure. Note
that this approach is more robust in detecting the border nodes than
the techniques based just on the geographical location of the nodes
(e.g., near the surface or at the periphery), as it considers
simultaneously the structural and dynamical properties of the network,
provided by the diversity entropy. In addition, the present approach
is particularly useful when applied to non-geographical networks,
where the identification of the network border is not possible.

\textbf{(ii) Finding the kernel nodes:} This step consists in
selecting nodes of the network based on their topological distance to
the border nodes, i.e., its \emph{kernel nodes}. A node is considered
a kernel node if its shortest path to any of the border nodes are
larger than some pre-defined threshold value. The detection of these
nodes can be performed by dilating all the border nodes --- the
dilation of a subnetwork (e.g.~\cite{Costa2006Generalized}) involves
incorporating all the immediate neighbors of that subnetwork --- and
selecting the nodes not reached by such a
dilatation. Figure~\ref{fig:ex_border_sel}(c) shows, in black, the
kernel nodes for the considered example.  Overall, note that this
border elimination approach is particularly efficient, as it
guarantees a safety distance between the border nodes and the kernel
nodes. The nodes lying in this intermediary region, i.e., between the
border nodes and the kernel nodes, are called henceforth
\emph{buffer nodes} (gray nodes in
Figure~\ref{fig:ex_border_sel}(c)). Note that the above procedure
guarantees that the application of diversity entropy to the analysis
of finite networks will be unaffected by border effects.

In order to further illustrate the concepts of the proposed
methodology, Figure~\ref{fig:ex_border_sel}(d) shows a diagram that
represents the relationship between border, buffer and kernel nodes.

\section{Applications}
In order to illustrate the efficiency and robustness of our proposed
method, we applied the border detection and diversity entropy analysis
to two real-world networks of different types.  The first network
considered is the bone vascular complex network, a strictly
geographical network, whose border nodes are know a priori. The second
example studied is the US air lines network, which is a small-world
network whose border nodes were not known and needed to be estimated.

\subsection{Bone vascular complex network}

\begin{figure}[ht]
\begin{center}
    \includegraphics[width=130mm]{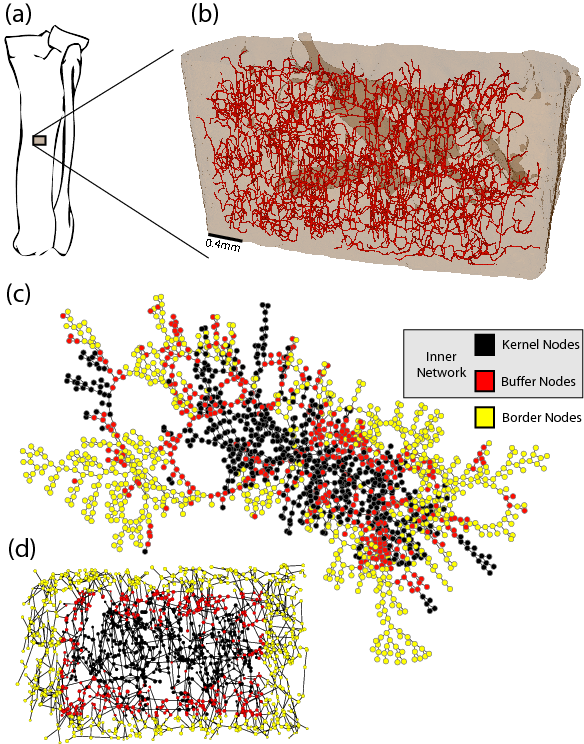}
    \caption{\textit{Topological and geographical view of the bone
    network.} (a) Posterior view of the pig tibia. The boxed delimited
    the scanned region. The position is at half the tibial length at
    the posterior side of the tibia. (b) Canals network embedded in the
    bone surface.  (c) Topological view of the bone network. The
    border nodes are shown in white (yellow), the buffer nodes are
    shown in gray (red) and the kernel nodes are shown in black. The
    inner network is composed by the buffer and the border nodes. (d)
    2D projection of the three-dimensional bone network and the
    respective border, buffer and inner nodes. Images (c) and (d)
    built using the software Cytoscape -
    \textit{http://www.cytoscape.org}.}  \label{fig:network}
\end{center}
\end{figure}

The first example studied in this work was a bone network, responsible
for conveying blood vessels through the tibia bone matrix. The
structural and dynamical aspects of this network are important to
understand the blood irrigation of the bone as well as in helping the
development of clinical procedures.

In order to obtain the bone network, a $\mu$CT sample of a right
tibia was obtained from a pig at 6 weeks of age. At six weeks, the
skeleton is still in a development stage because a mature structure
is obtained only after 100 weeks approximately. The bone specimen of
4 x 4 x 4 mm was extracted from the posterior cortex at diaphysis
level, as showed in the figure~\ref{fig:network}(a) (for more
details about the dataset obtention please refer
to~\cite{Tanck2006Bone}).

Next, the bone specimen was scanned into an image dataset which was
subsequently used to obtain the three-dimensional representation of
the bone shown in figure~\ref{fig:network}(b). In this figure, the
gray points corresponds to the bounding surface of the cortical
portion. Our main focus of interest in this paper are the canals
represented by the lines in figure~\ref{fig:network}(b), which are
the pores within the cortex. Is important observe that the canals
are not represented with their real thickness. The whole canal
structure was obtained after the image was submitted to a thinning
algorithm in order to reduce the redundance of the structure. Our
next step was to map the canal structure into a graph. In order to
do so, we developed an algorithm that proceeds through the thinned
canals and marks the bifurcation points. These bifurcation points
were mapped as the nodes in the graph representation, while the
canals were considered as the edges. A topological view of the final
graph can be seen in Figure~\ref{fig:network}(c) and the respective
2D projection of the three-dimensional network is shown in
figure~\ref{fig:network}(d). The obtained network has 1800 nodes and
2131 edges, with a mean degree of 2.4. This value is in agreement
with that obtained in a previous study with a femur of an adult cat
\cite{Viana2006Bone}. It should be noticed that though other papers
have used the 3D techniques and thinning procedure to study the
internal organization of the trabecular and cortical bones
\cite{Petersson2006ThinningBone, Cooper2003ThinningBone,
Xie2003ThinningBone, Saha2000ThinningBone}, very few articles have
used complex networks concepts in order to study these
systems~\cite{Viana2006Bone}.

A problem with geographical networks is that very often we do not have
information about the whole system, but only some parts sliced from
the whole. In these cases, non-local measurements suffer influences of
the border due to finite size effects. Such effects are verified in
the bone network because it corresponds to only a portion of the
original bone. Because the portion of bone under study was sliced from
the original structure, we know which of the nodes belong to the
border (i.e. they are the nodes adjacent to the sliced surfaces).
Therefore, it is important to identify the set of nodes unaffected by
the borders of the network, in order to constrain the respective
analysis at those nodes.

\subsection{US air lines}

The second network considered in this work is the US traffic air
system. In this case, the nodes of network represent the airports
and two nodes are connected if there is a flight between them. We
study a unweigthed version of the original network available at
Pajek website
(http://vlado.fmf.uni-lj.si/pub/networks/data/default.htm). The
network has 332 nodes and 2126 edges given a average degree equal
12.8. Although the nodes has a very well established Euclidian
coordinates, as the bone network, the US airline network appear a
small-world behavior~\cite{Barabasi2007} due to non local
connections capable to link very distant nodes.  It is particularly
difficult to identify, a priori, the border nodes in this network
because they do not correspond necessarily to the airports at the
geographical border (such airports can still serve many flights).

\section{Results}

Considering the bone vascular network and its respective complex
network shown in Figure~\ref{fig:network}, the diversity entropy of
each node was estimated by using the deterministic and fully
accurate algorithm outlined in~\cite{Costa2008Superedges}.
Figure~\ref{fig:network}(c-d) shows, in white (yellow), the selected
border nodes of the bone network identified by considering a small
region near the lateral surfaces of the bone. The buffer nodes are
shown in gray (red) and the kernel nodes in black. In this analysis,
the kernel nodes are at a minimum of three edges of distance from
the border nodes.

\begin{figure}[ht]
\begin{center}
\includegraphics[width=130mm]{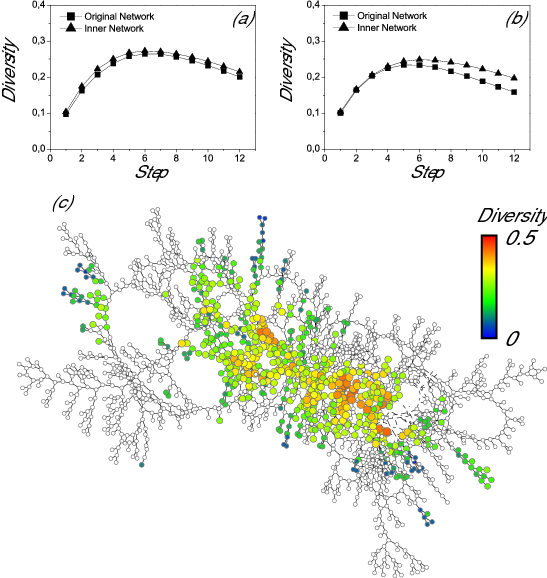}
\caption{Diversity entropy of bone vascular network.  (a) Comparison
of the diversity entropy considering only the kernel nodes of the
original network and the inner network. In this graph the mean
diversity is shown along different steps. Note that the results are
quite similar. (b) When the same comparison is taken considering all
the nodes, i.e., the whole original network and the whole inner
network, the similarities disappear. This fact is, in part, a
consequence of the inclusion of the border nodes in the analysis, and
it reinforces the fact that the selected kernel nodes used in (a)
are not influenced by border effects. (c) Diversity entropies of the
kernel nodes for the original network at step $h=4$.}
\label{fig:diversity}
\end{center}
\end{figure}

In order to show that the selected kernel nodes are not influenced
by the border nodes, we defined a new network, named as \emph{inner
network}, which is formed only by the buffer and the kernel nodes of
the original network (see figure~\ref{fig:network}c). The diversity
entropy of the inner network was computed and compared with the
original network. The graph presented in
Figure~\ref{fig:diversity}(a) shows the mean diversity entropy
considering only the kernel nodes, for both networks and for $h=1$
to $12$ (the maximum number of steps considered in our present
analysis). It is clear from this graph that the kernel nodes are not
affected by the border nodes, as no considerable differences can be
detected in their diversity mean, even for high values of $h$.
Contrariwise, when all the nodes of both networks are considered,
without any buffer region, the differences in the mean diversity are
substantial, as illustrated in figure~\ref{fig:diversity}(b),
reflecting intense border effects.

The diversity values for each kernel node considering $h=4$ are
shown in figure~\ref{fig:diversity}(c). Observe the cluster of nodes
with high diversity values in a specific region of the network. This
region probably corresponds to distribution center (for blood and
other substances), as its provides access to a considerable number
of nodes of the network in a uniform fashion, while the nodes with
low diversity enhances the access to a smaller number of nodes of
the network and can be considered as a destination region.

\begin{figure}[ht]
\begin{center}
\includegraphics[width=80mm]{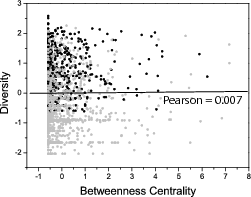}
\caption{\textit{Betweenness Centrality versus Diversity Entropy}.
The scatterplot shows the value of betweenness centrality against
the diversity entropy for each node in the bone network. The
scattering revels no direct correlation between both measurements,
as also confirmed by the low Pearson correlation coefficient
(0.007). The black color is related to the kernel nodes.  The
remaining nodes are shown in grey. The betweenness centrality was
calculated by using the NetworkAnalyzer Plugin for Cytoscape
\textit{http://med.bioinf.mpi-inf.mpg.de/netanalyzer/index.php}.}
\label{fig:centrality}
\end{center}
\end{figure}

Also, in order to show that the selected kernel nodes are not
necessarily the most central ones (in the sense of betweeness
centrality -- the number of times that a node is crossed by shortest
paths between all the other nodes~\cite{Brandes2001Centrality}), a
scatterplot is depicted in Figure~\ref{fig:centrality} showing the
diversity entropy against the betweenness centrality for all nodes.
It is clear from this result that the diversity entropy and betweeness
centrality are almost completely uncorrelated (Pearson = 0.007),
confirming that the diversity entropy provides important complementary
information about the structure and dynamics of the analyzed network.

\begin{figure}[ht]
\begin{center}
\includegraphics[width=130mm]{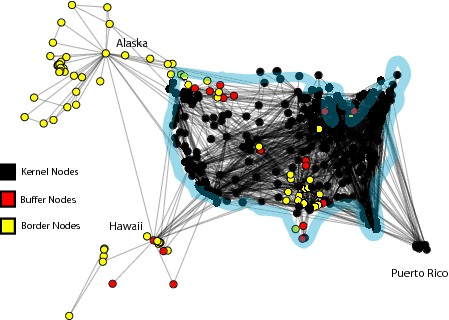}
\caption{US airlines network. The border detection methodology
identified the border nodes (yellow/white), the buffer nodes
(red/gray) and the kernel nodes (black).} \label{fig:airport1}
\end{center}
\end{figure}

The results of the application of the proposed methodology in the
analysis of the diversity of the US air lines network is shown in
Fig.~\ref{fig:airport1}. In this figure, the border nodes, shown in
white (yellow), were obtained considering nodes whose diversity were
lower than 90\% of the maximum diversity value for $h=4$. The
detection of the buffer nodes was slightly modified. As the US air
lines network is a small world network, there are great chances that
the dilatations of the border nodes reach the hubs of the network. As
a consequence, the hubs of this network would hardly be considered as
kernel nodes. In order to avoid such an effect, the following
restriction was included in the dilation process: the incorporation of
a node occurs only if a percentage of the neighbors of the respective
node are border nodes.

\begin{figure}[ht]
\begin{center}
\includegraphics[width=130mm]{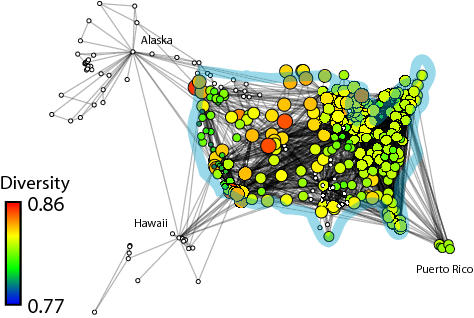}
\caption{Diversity entropy for the kernel nodes of the US air lines
network.} \label{fig:airport2}
\end{center}
\end{figure}

In the case of the US air lines network, a node was included in the
dilatation if at least 30\% of its neighbors were border nodes and
if its shortest path to any of the border node had a maximum
distance of three edges. The result of application of this procedure
is shown in Figure~\ref{fig:airport1}, where the buffer nodes are
shown in gray (red) and the kernel nodes are shown in black. In
addition, the diversity entropy of the kernel nodes for $h=4$ is
shown in Figure~\ref{fig:airport2}.

An interesting result is observed in the distribution of the kernel
nodes. All of them are situated at the main territory of the US,
except for the airports located at Puerto Rico. This means that the
periphery of this network, i.e., the nodes that was classified as
border or buffer, are composed mainly by the nodes representing the
insular territories of US and the state of Alaska.  The Puerto Rican
airports were classified as kernel nodes because they serve a large
quantity of routes.  The diversity entropy values obtained for the
kernel nodes revealed that American airports at the west coast tend to
allow more effective and uniform access to a larger number of
airports than those at the east coast.

\section{Concluding Remarks}

Frequently, a complex networks being investigated represents only a
sliced and/or sampled portion of a larger structure which is not
available or which would otherwise be too large for the analysis.
Such networks necessarily imply the existence of \emph{border nodes},
i.e. those nodes connecting the analysed structure to the remainder of
the larger original structure.  Because such border nodes imply in
strong bias on several measurements of the topology of the networks
under analysis, they need to be identified and avoided. Indeed, in
several cases it is not only the border nodes which need to be
avoided, but also their neighbors, which we have called \emph{buffer
nodes}.  Two situations arise in practice: the border nodes are
already known or have to be identified.  The present work reported an
objective and sound approach for identification of border nodes by
using a recently introduced measurement, namely the diversity
entropy~\cite{Costa2008Diversity}.  In addition, we also show how the
effects of such border nodes on the calculation of the diversity
entropy measurement can be minimized by considering a buffer zone
obtained by dilating the border nodes.

Once the border nodes have been identified, the \emph{kernel nodes}
are found by considering the topological distance from the border
nodes. The kernel nodes are then considered for the measurements and
analyses of the network properties.  The effectiveness of the
suggested methodology was corroborated with respect to the diversity
entropy analysis of a network of canals extracted from a sliced
portion of bone. Because of technical limitations, it is virtually
impossible to obtain the structure corresponding to the whole original
bone (tibia), so that samples have to be sliced from it, implying in
severe border effects.  The obtained results indicated a large
dispersion of diversity entropy values amongst the nodes in the
considered network, revealing that different nodes have distinct
topological and dynamical interactions with the other portions of the
bone canals system, with the most central nodes (i.e. those with
higher diversity entropy) being capable of effectively reaching a
larger number of other nodes. In order to illustrate the
identification of border nodes, we considered the US air
transportation system.  The identification of the border nodes
involved the calculation of the diversity entropy for all the nodes in
the given network and taking the nodes with smaller diversity entropy
values as borders.  The consideration of the buffer zone then allowed
the calculation of meaningful diversity entropies for the kernel
nodes.

It is important to keep in mind that the reported framework is by no
means limited to geographical networks, being immediately applicable
to any type of network.  In addition, the method can also be applied
for the elimination of border effects while calculating several other
topological measurements.

\noindent{}\textbf{Acknowledgments:} Matheus P. Viana thanks FAPESP
(07/50882-9) for financial support. B. A. N. Traven\c{c}olo is
grateful to FAPESP (07/02938-5) for financial support. Luciano da F.
Costa thanks FAPESP (05/00587-5) and CNPq (301303/06-1) for
sponsorship. Esther Tanck thanks NWO-STW (NPG.6778) for financial
support.

\section*{References}
\bibliographystyle{unsrt}
\bibliography{pigBoneAccess}

\end{document}